\documentclass[preprint,eqsecnum,aps,showpacs]{revtex4}
\usepackage{dcolumn}
\usepackage{graphicx}
\usepackage{latexsym}

\begin{document}

\title{Braneworlds in six dimensions: new models with 
bulk scalars} 

\author{Ratna Koley \footnote{Electronic address : 
{\em ratna@cts.iitkgp.ernet.in}} ${}^{}$ 
and 
Sayan Kar\footnote{Electronic address: {\em sayan@cts.iitkgp.ernet.in}}${}^{}$}
\affiliation{Department of Physics and Centre for Theoretical Studies  \\
Indian Institute of Technology, Kharagpur 721 302, India}

\begin{abstract}
Six dimensional bulk spacetimes  with 3-- and 4--branes are constructed
using certain non--conventional bulk scalars as sources. In particular, we 
investigate the consequences of having the phantom (negative kinetic energy) 
and the Brans--Dicke scalar in the bulk while obtaining such solutions. 
We find geometries with 4--branes
with a compact on--brane dimension (hybrid compactification)
which may be assumed to be small 
in order to realize a 3--brane world.
On the other hand, we also construct, with similar sources, bulk
spacetimes where a 3--brane is located at a conical 
singularity.
Furthermore, we investigate the issue of localization of matter fields (scalar,
fermion, graviton, vector) on these
3-- and 4--branes and conclude with comments on our six dimensional 
models. 
\end{abstract}


\maketitle

\section{Introduction}



The idea that our world could be visualized as a co-dimension two brane
in six dimensions was first noticed by 
Akama \cite{akama} way back in 1982. Using the dynamics of the Nielson-Oleson vortex 
solution of the Abelian Higgs model in six dimensions, he was able to
localize our spacetime within a 3-brane, with Einstein gravity being
induced through the fluctuations of the brane.
Subsequent to the work of Akama, Rubakov and Shaposhnikov {\cite{rubakov}} 
and Visser {\cite{visser} pursued closely related ideas around the same
time. These were, by and large, the early ideas on the notion of 
{\em alternatives
to the usual Kaluza-Klein compactification}, which are being hotly pursued 
today following the work on large extra dimensions {\cite{add1}} and 
warped extra dimensions {\cite{rs2}}.

This article is focused on braneworld models with codimension greater
than one. In particular, we shall be exclusively concerned with bulk
spacetimes in six dimensions generically represented by the line element: 

\begin{equation}
ds^2 = \sigma(x^{a}) g_{\mu\nu}dx^{\mu}dx^{\nu} +
\gamma_{ab}(x^{a})dx^{a} dx^{b}  \label{eq:genmetric}
\end{equation}

where $\sigma(x^{a})$ is a conformal factor depending on the extra
coordinates and $\gamma_{ab}(x^{a})dx^{a} dx^{b}$ is 
line element representing the extra dimensional part (for a 6D bulk
this is two dimensional). 
Certain specific solutions of the 6D Einstein equations  
with a positive bulk cosmological constant had been obtained 
in \cite{rubakov} where the extra dimensions are non-compact and assumed to
be unobservable at low energies. 
Following the recent string inspired phenomenological brane world
models proposed by Randall and Sundrum \cite{rs2}, 
a fair amount of activity has been generated involving possible
extensions and generalizations, among which, co--dimension two models
in six dimensions have been a topic of increasing interest.  


Let us now briefly review the work done on codimension two models in
six dimensions. 
Shortly after the work of Randall--Sundrum involving a warped geometry, 
several proposals came up which made use of two extra dimensions
{\cite{2extra}}. Most of these articles followed the original viewpoint 
(Akama, Rubakov and Shaposhnikov) with
the four dimensional world being a cosmic--string like topological defect
\cite{gherghetta}
as opposed to the domain wall type defect in codimension one models.
Chodos and Poppitz \cite{2extra} first brought in the idea
of warped braneworlds of codimension two where the presence of a brane 
appeared through the existence of a location with a conical deficit in the  
bulk. 
In particular, for some cases, the resolution of the hierarchy problem was also
achieved and unlike Randall--Sundrum, there did not exist any
fine--tuning between the brane tension and the bulk cosmological
constant. Gravity was also found to be localisable on the defect
(eg. in \cite{gherghetta}).
A useful review on topological defects in higher dimensional models 
and its relation to braneworlds is available in {\cite{roessl}}.

Furthermore, Leblond {\em {et.al.}}  
obtained a set of consistency conditions for braneworld scenarios with a
spatially periodic internal space in \cite{sumrule}, from which one can 
see that the necessity of a negative tension brane appears in five
dimensions and is absent in the higher dimensional
constructions. It is also apparent from the
multigravity scenario discussed in \cite{kogan} that the radion
stabilization problem and the presence of a negative tension brane is an
artefact of five dimensional spacetime. The non-trivial curvature of
the internal space in the case of two or more extra dimensions provides
the necessary bounce configuration of the warp factor without the need
of any negative tension brane \cite{6dmodel}. Generically, in six dimensions,
two types of constructions are around. One of them involves 4-branes 
which localize gravity but one of their dimensions is compact, 
unwarped and of Planck length \cite{kogan}. The
other type of construction has conical singularities
which support 3-branes (these can be
of positive, negative and zero tension depending on the angle deficit,
angle excess or no angle deficit respectively) \cite{gherghetta}. 

From the above introduction, it is evident that in the context of 
factorisable (un-warped) as well as warped bulk spacetimes
a fair amount of work has been carried out in the recent past. 
Apart from model construction, the question of
solving the cosmological constant problem has been the primary issue
addressed in several articles \cite{cosmoconstant6d}. Other aspects
such as cosmology, brane gravity etc. have been discussed by 
numerous authors {\cite{codim2gen}.
A list of some recent articles on codimension two models is provided in
\cite{codim2recent}.


It is well--known by now that in the braneworld scenario it is necessary 
to introduce dynamics which can determine the location of the branes
in the bulk. 
Ever since Goldberger and
Wise \cite{goldwise} added a bulk scalar field to fix the location of the
branes in five dimensions, investigations  
with bulk fields became an active area of research.  
The consequences of different types of bulk scalars in the bulk 
spacetime geometry and their phenomenological implications have been 
looked into in great detail over the last few years. 
For example, in the RS-II \cite{rs2} set--up, it has been noted that
spin half fields cannot be localized on the brane by
the gravitational interaction only \cite{bajc,loc5}. Thus, it becomes
necessary 
to introduce additional non-gravitational interactions (eg. a fermion--scalar
Yukawa coupling, say) to get spinor fields confined to the brane. 
A simple choice of such a non--gravitational field in the bulk is a scalar 
field coupled to gravity \cite{loc5}. 



Motivated by the above--mentioned need of bulk scalars,
we carry out our search for novel bulk spacetimes in six dimensions with such 
bulk scalars of various types as the sources in the six dimensional Einstein
field equations. The advantage of studying various
types of models is  related to the fact that it helps in revealing 
a wider spectrum of possibilities. 
In recent times, some interesting solutions of new brane
models in 6D have been obtained. For example,  
a general regular warped solution with 4D Minkowski spacetime in 
six-dimensional gauged supergravity is obtained in \cite{6dnew}.    
A simple exact solution of 6-dimensional braneworld which 
captures some essential features of warped flux compactification, 
including a warped geometry, compactification, a magnetic flux, 
and one or two 3-brane(s) is found in \cite{6dnew}. 
Higher dimensional fermions in a non-singular 6D brane background 
with an increasing warp factor has also been studied \cite{6dnew}. 
It has also been claimed in \cite{6dloc} that all the zero 
modes of the standard model fields can be localized on a 
single brane by means of only the gravitational interaction. 

In  our first example here, we investigate the effect of having a 
phantom field  in the six dimensional bulk spacetime.
Recently, in cosmology, the phantom scalar has
been widely used \cite{phantom} to explain dark energy and the 
accelerated expansion of the universe. The phantom is a hypothetical 
scalar field with a wrong-sign (negative) kinetic energy term in its 
Lagrangian. Even though questions of stability (unbounded negative energy) 
arise in such models, phenomenologically (eg. in cosmology), 
they have been useful in explaining various scenarios. In order to
justify the existence of the phantom, a model of phantom energy has been 
constructed in {\cite{singleton}}, using the graded
super Lie algebra SU(2/1), where the negative kinetic energy term seems to
arise naturally. Furthermore, in our earlier work, we have seen that 
the presence of such a scalar field in the 5D bulk plays
a crucial role in localizing massless as well as massive fermions on
the brane \cite{rksk2}. 
Our investigation here is based on the exact solutions of the full 6D 
Einstein-phantom scalar equations. In the exact background
geometry obtained from this setup it is possible to have the zero modes 
of all the standard model fields and gravity to be localized on 4--branes.
 
In our next example, we look for the consequence of introducing a Brans-Dicke
scalar in the bulk (i.e. consider 6D Brans--Dicke gravity).
Recently, the role of a Brans--Dicke scalar in five dimensions and the
corresponding bulk solutions have been investigated in \cite{mikhailov}.
We obtain here, the bulk solution and the conditions for confinement of
gravity as well as other matter fields on the brane. 

In the above two models, the
extra dimensional space is of finite volume with a negative
Ricci scalar (hyperbolic two dimensional geometries). Additionally, the
models with a phantom or a Brans-Dicke bulk scalar both involve four brane
constructions with an assumed,
on--brane, compact, extra dimension (hybrid compactification).  

Finally, we concentrate on models where a brane embedded in a six 
dimensional bulk is  realized via a conical deficit in the bulk spacetime 
(similar to the 
topological defect type braneworld models). We also investigate the issue
of localization of fields for this class of models and comment on 
conditions under which localization of all fields is possible for a
sufficiently broad class of warp and extra dimensional factors.
  
The organization of our paper is as follows. In Section II, we have obtained
the exact solutions of the Einstein-scalar equations for the bulk phantom and
the Brans-Dicke scalar.  
We start with codimension two branes and finally generalize the results in
higher codimensions in some particular cases. Section III discusses 
localization of gravity and other matter fields on the brane through the
existence of normalizable  zero modes on the brane. 
Section IV deals with models with a conical deficit at the brane location
and the issue of localization in such examples. 
In the last section, we conclude with discussions and open issues.


\section{The ansatz, equations of motion and solutions for
  non-singular 4-branes}

\subsection{The models and the bulk solutions}

Let us begin with the most general metric ansatz for a warped
brane embedded in six dimensions obeying four dimensional Poincare
invariance: 

\begin{equation}
ds^2 = g_{MN} dx^{M} dx^{N} = e^{2f(r)} \eta_{\mu\nu}dx^{\mu}dx^{\nu} + d r^2 +
e^{2 g(r)} L^2 d\theta^2  \label{eq:metric}
\end{equation}

where the radial coordinate $r$ is infinitely extended ($0<r<\infty$) 
and the compact
coordinate $\theta$ ranges from $0 \leq \theta \leq 2\pi$. L is 
additional parameter characterizing the extra compact direction
on the 4--brane. We also assume that
the warp factors are functions of the extra dimensional radial coordinate, $r$,
only.  

We will now focus on the time-independent solutions of the Einstein equations 
for two types of bulk scalar field sources -- (i) a phantom scalar 
and (ii) a Brans-Dicke scalar. In this context an obvious 
question occurs -- why do we choose such fields in the bulk? 
The answer lies in our ignorance about what could be there in the
bulk. With vacuum, we do not find appropriate solutions and with
a negative $\Lambda$ we have the Randall--Sundrum type models.
In search of further models with distinct characteristics
we consider as a first simple choice : scalar fields in the bulk.
The usual scalar field (with a potential) does not seem to provide a
useful (and non--trivial) solution--hence we turn to non--standard scalars 
such as the ones mentioned above. Honestly speaking, there is no rationale 
about our choice of matter in the bulk. However,   
if consequences which result on the brane are physically relevant 
then we might call our choice as reasonable, with the model being capable
of representing our usual four dimensional world as a surface in
the bulk. 

\subsubsection{Model-I : Phantom scalar field in the bulk}

The six dimensional action for a bulk scalar field
(phantom : with a ``wrong'' sign kinetic term) in a potential V($\phi$), 
minimally coupled to gravity in the presence of a cosmological constant 
is given by   

\begin{equation}
S = \int   \sqrt{-{}^{(6)}g}\left[ \left( R - 2 \Lambda\right) \frac{M^4}{2} +
  \frac{1}{2} g^{AB} \nabla_{A} \phi \nabla_{B} \phi - V(\phi)\right]
 d^{6}x \label{eq:action}
\end{equation} 

where, M corresponds to the six dimensional fundamental mass scale. We assume
henceforth that 
the scalar field is a function of the coordinate r only.   
Variation of the action (\ref{eq:action}) with respect to the metric 
and the scalar field leads to the following field equations for the 
Einstein-scalar system :

\begin{eqnarray}
6 f'^2 + 3 f'' + 3 f' g' + g'' + g'^2 & = &  \alpha
({\phi'^{2}}/{2} - V)  - \Lambda \\ 
 & = & 10 f'^2 + 4f'' \\
6 f'^2 + 4 f' g'   & = & \alpha
(- {\phi'^{2}}/{2} - V ) - \Lambda \\ 
(\phi'' + 4 f'\phi' + g'\phi') & = & - \frac{\partial V}{\partial \phi}
\end{eqnarray}

where $\alpha = \frac{1}{M^4}$ and the prime denotes a derivative with 
respect to $r$. Note that the
scalar field equation is not independent, it can be obtained from the
other three equations. To obtain an exact analytical solution for
the warp factors and the scalar field we first work with $V(\phi) = 0$
and also tune the bulk cosmological constant to be zero. Further
, assuming the radius of the compact dimension to be of the order of Planck scale, we obtain
the following solutions for the warp factors and the bulk scalar
field as

\begin{eqnarray}
e^{2f(r)} & = & e^{\frac{k}{2} r} \hspace{.2cm} \mbox{and}  \hspace{.2cm} 
e^{2g(r)}  =  e^{- 2 k r} \label{eq:warp2}\\
\phi(r) & = & \left (\frac{5 k^2}{4 \alpha} \right)^{\frac{1}{2}} r \label{eq:scalar}
\end{eqnarray}

where, $ k$ is an arbitrary constant. 
Note the distinct nature of the warp factors - the brane part 
is a growing function of$ r$ and the other part is a decaying function. 
The geometry of the spacetime has a Ricci curvature which can be
obtained from the formula :

\begin{equation}
{}^{(6)}R=\left (-20 {f'}^2-8f''\right ) + \left (-2g''-{2g'}^2 \right )
-8f'g'
\end{equation}

With the warp factors for the case of the phantom scalar field one
obtains ${}^{(6)}R=-\frac{5}{4}k^2$. The two dimensional extra dimensional
space has ${}^{(2)}R=-2 k^2$. The volume of the two dimensional
piece is also finite and is given by $\frac{2\pi L}{k}$.  
Thus, both the full space
and the extra dimensional space (considered separately)
are both of negative Ricci curvature (AdS) with the latter having 
a finite volume.
 
The results are different in nature from those obtained 
by considering 
only a cosmological constant in the bulk \cite{kogan} and for a smooth 
local defect in the bulk \cite{gherghetta}. The energy-momentum tensor 
for the bulk field which gives rise to the above solution 
has components given by 


\begin{eqnarray}
\rho & = & - p_{x,y,z}  =  - \frac{5}{8} k^2  \label{eq:em1} \\
p_{r} & = & - \frac{5}{8} k^2 \hspace{.2cm} \mbox{and} \hspace{.2cm}
p_{\theta}  = \frac{5}{8} k^2 \label{eq:em2}
\end{eqnarray}


It is clear from the above expressions that the matter stress energy 
which gives rise to the background geometry (\ref{eq:warp2})  
violates all the energy conditions ( namely WEC, SEC, NEC ) \cite{visserbook}. 
However,
the energy density and pressures are constant and therefore bounded (and,
obviously less problematic than an infinitely unbounded negative
energy density). 
Thus, our solution, as far as the bulk source is concerned, 
is not very different from the usual Randall--Sundrum solution 
in five dimensions, where the
bulk has only a negative cosmological constant. In the phantom case, however,
we have $p_\theta= - \rho $,  $p_r= \rho$ (unlike the cosmological constant which
would have required $p_i = -\rho$ for all $i$). 

We can generalize the results for situations where the total 
spacetime dimensions are more than six. 
The extra dimensional space is constructed with one non-compact dimension 
and ($p$ - 1) compact dimensions 
where $p$ is the number of extra dimensions.  The warp factors and 
the scalar field are assumed to be a function
of only the radial dimension $r$. 
The exact analytical solution of the Einstein-scalar
equations in (4+$p$) dimensions follows turns out to be :

\begin{equation}
ds^2 =  e^{(p-1)\frac{k r}{2}} \eta_{\mu\nu}dx^{\mu}dx^{\nu} + d r^2 +
e^{- 2kr} \eta_{mn} dy^m  dy^n \label{eq:metricp}
\end{equation}

where m,n runs from 1 to $p$.
The nature of the warp factors remain same as those obtained in six dimensions. 
In this case also the scalar field is a monotonically increasing function 
of $r$ as given in equation (\ref{eq:scalar}).

\subsubsection{Model-II : Brans-Dicke scalar field in the bulk}

We now introduce another example of a bulk scalar 
coupling with gravity ( namely the Brans-Dicke coupling) in a six dimensional
spacetime. The scalar-tensor Brans-Dicke theory \cite{bdick} is known to be an 
alternative theory to Einstein's  
general relativity. It is similar to GR, except the reciprocal 
of the gravitational constant is itself a one-component field, 
the scalar field $\phi$, which is generated by matter through 
an additional equation (the scalar field equation). 
Thus $\phi$ as well as usual matter both play their 
roles in determining the metric via a modified version of Einstein's 
equations. 
In fact, Brans-Dicke theory is distinguishable from general relativity 
only by the value of its single dimensionless parameter $\omega$ which 
determines the effectiveness of matter in producing $\phi$. The larger 
$\omega$, the closer the Brans-Dicke theory predictions are to those for 
general relativity.
Initially a popular alternative to General Relativity, the Brans-Dicke 
theory lost favor as it became clear that $\omega$ must be very large
- an artificial requirement. Nevertheless, the theory has 
remained a paradigm for the introduction of scalar fields into 
gravitational theory, and as such has enjoyed a revival in 
connection with low energy effective theories derivable from
quantum string theory. The so--called dilaton gravity can be 
identified as a $\omega=-1$ Brans Dicke theory.  

In our work here we examine the consequences of having a Brans--Dicke scalar
in the bulk. Apart from obtaining the line elements satisfying the equations
of motion we also study the localization
of fields on the brane in these models. 

For Brans-Dicke theory the gravitational and scalar field equations are
given as: 

\begin{eqnarray}
G_{AB} & = & \frac{8 \pi}{\phi} {T_{M}}_{AB} + \frac{\omega}{\phi^2}
\left( \nabla_{A} \phi \nabla_{B} \phi - \frac{1}{2} g_{AB}
\nabla_{R} \phi \nabla^{R} \phi \right) + \frac{1}{\phi} \left(
\nabla_{A} \nabla_{B} \phi - g_{AB} \Box^2 \phi \right) \\
\Box^2 \phi & = & \frac{8 \pi}{3 + 2 \omega} {T_{M}}_{A}^{A}
\end{eqnarray}

We choose, the energy-momentum tensor for matter to 
to be zero {\em i.e.} ~ ${T_{M}}^{AB} = 0 $. With this choice we 
shall now find exact analytical solutions for the background geometry first
in six dimensions then in arbitrary dimensions. 

In the background geometry defined by the line element in equation 
(~\ref{eq:metric}), the above equations reduce to 
\begin{eqnarray}
6 f'^2 + 3 f'' + 3 f' g' + g'' + g'^2 &=&  \left [- \frac{\omega}{2} \frac{\phi'^2}{\phi^2} -
  \frac{1}{\phi}(\phi'' + 3 f' \phi' + g' \phi') \right] \\
6 f'^2 + 4 f' g'  &=& \left [ \frac{\omega}{2}\frac{\phi'^2}{\phi^2} - \frac{1}{\phi} (4
f' \phi' + g' \phi')\right] \\
10 f'^2 + 4f'' &=& -\left [\frac{\omega}{2} \frac{\phi'^2}{\phi^2} +
  \frac{1}{\phi}(\phi'' + 4 f' \phi') \right] \\
 \phi'' &+& 4f' \phi' + g' \phi' = 0
\end{eqnarray} 

As before, the scalar field is assumed to be a function of fifth
coordinate $r$ only and the prime denotes the derivative with respect to
this coordinate. An exact solution of the above set of equations is given as:
\begin{eqnarray}
f(r) & = & k_{1} r \hspace{.5cm} \mbox{and}  \hspace{.5cm}
g(r)  =  k_{2} r \label{eq:warp4}\\
\phi(r) & = & e^{k_{3} r} \label{eq:scalar2}
\end{eqnarray}

where the constants are constrained by the following relations 
\begin{eqnarray}
4 k_{1} + k_{2} & = & - k_{3}\\
\frac{\omega}{2}k_3^2  =  k_2 k_3 & - & 10 k_1^2 
= 6k_1^2+4k_1k_2-k_3^2=-6k_1^2-3k_1k_2-k_2^2+k_1k_3 \nonumber
\end{eqnarray}

Using the first constraint in the second one arrives at a equation involving
$k_1$ and $k_2$ which can be solved to give :

\begin{equation}
\frac{k_1}{k_2}=\left [ \frac{1-\vert\omega\vert}{5-4 {\vert\omega\vert}}\pm 
\frac{\sqrt{-6+5\vert{\omega}\vert}}{2(5-4\vert{\omega}\vert)}
\right ] 
\end{equation}

where $\omega=-\vert{\omega}\vert$. Also $\vert{\omega}\vert>\frac{5}{4}$ and 
$\omega$ is necessarily negative.
One can further show that:
\begin{eqnarray}
\omega = - \frac{4 k_{1}^2 + k_{2}^2 + k_{3}^2}{ k_{3}^2}
\end{eqnarray}

Using the above constraints one may construct a typical example. 
If $k_1=k(>0)$ and $k_2=-k$
then $k_3=-3k$ and $\omega=-\frac{14}{9}$. 
Thus the brane has growing warp factor, the scalar field decays 
for larger values of the extra dimension and the
extra dimensional geometry (two dimensional) has a decaying exponential
(anti-de Sitter in two dimensions). On the other hand, we may also have
$k_1=-k$, $k_2=k(>0)$ for which $k_3=3k$ and $\omega =-\frac{14}{9}$.
This gives a decaying brane warp factor, a growing scalar field and a
de Sitter extra dimensional space. 
Many other possibilities exist with varying values of $k_1$, $k_2$,
$k_3$ and $\omega$. A plot of $\frac{k_1}{k_2}$ as a function of
$\vert{\omega}\vert$ (Fig. 1) illustrates the allowed range of solutions.

\begin{figure}
\includegraphics[width= 9cm,height=5.6cm]{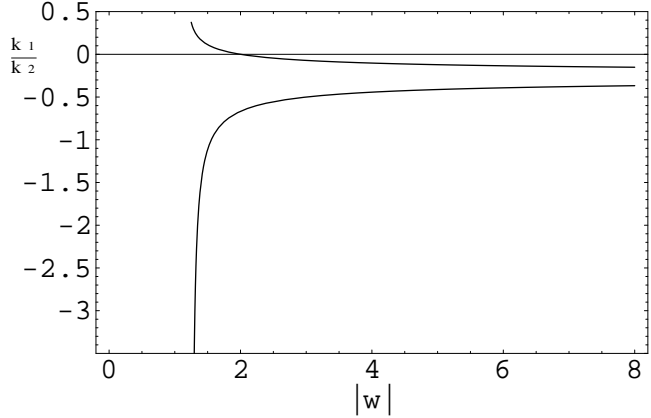}
\caption{The ratio $k_1/k_2$ as a function of $\vert \omega \vert$. The upper
(lower) graphs are the ones for the $-$($+$) sign before the square root } 
\end{figure}


For a generalization of the solutions to co-dimension p branes we
derive the scalar-gravity equations in (4 + $p$) dimensions. An exact static
solution of the spacetime geometry is obtained for relations 
\begin{eqnarray}
4 k_{1} + (p-1) k_{2} + k_{3} = 0 \\
\omega = - \frac{4 k_{1}^2 + (p-1) k_{2}^2 + k_{3}^2}{ k_{3}^2}
\end{eqnarray}

The scalar field is the same exponential function of $r$. The
constant $k_{3}$ is rescaled by the parameter $p$. The whole
six dimensional spacetime has, as before, a negative curvature scalar. 

\subsubsection{Placing the single branes}

In all the above models one needs to place branes in order to have a
{\em braneworld}. Following standard methods, we now take into account
the effects of placing branes in the above bulk geometries. In particular,
we evaluate the brane tension as a function of the parameters which appear in
the expression for the bulk fields.

To include a single 4--brane here, we notice that we need to modify the
bulk action with a 4-brane contribution (world volume action). Correspondingly,
the Einstein field equations change and the effect is seen in the
$G_{00}$ and $G_{\mu\mu}$ ($\mu=1,2,3$) terms through the presence of
a Dirac delta function term $\lambda \delta (r)$ (where $\lambda$ is the
brane tension). To achieve a delta function in the RHS of the
Einstein equation we modify the warp factors by extending the domain of
$r$ to $-\infty < r < \infty$ and replacing $r$ by $\vert r \vert$.
The $G_{00}$ Einstein equation finally yields :

\begin{equation}
\lambda =3k_1+k_2 =-\frac{k}{4}
\end{equation}


\subsection{Localization of gravity and other matter fields} 

An important issue related to the viability of a braneworld model is
the question of localization of gravity and other matter fields on the
brane {\cite{bajc,loc5}}. 
To address this point, we now consider localization of different types of 
matter fields in the context of the models discussed above. 
 
In order to see whether the fields are confined or not we first employ
the simplest test, originally outlined in {\cite{bajc}}. 
For fields of different spins 
in six dimensions we assume, at the outset, 
that they are independent of the extra coordinates. The consistency
check is then done by showing that the effective coupling constants 
emerging after dimensional reduction are non-vanishing and finite.   
In a sense, this approach assumes localization as a starting point. 
On the other hand, one may consider the fields to be dependent on the 
extra dimensional
coordinates and then solve the relevant equations to see whether 
the behavior of the fields conforms with localization. 
To have localized modes one requires the
extra dimensional part of the field to peak around the brane and 
the full solution to be normalizable and finite everywhere. 

\subsubsection{Gravitational field (spin 2):}

We analyze the spectrum of linearized tensor fluctuations to see
whether gravity is localized on the brane so that the model remains 
consistent with the results of the usual Newtonian and 4D GR experiments. 
For a fluctuation of the full 6D metric $g_{MN} \rightarrow g_{MN} + h_{MN}$ 
one has a variety of polarizations, or graviton modes
\cite{6dmodel}. In six dimensions there are three kinds of modes - (i)
transverse traceless modes which are polarized along the Lorentz
invariant hypersurface represented by ${x^{\mu}}$, (ii) vector
modes polarized along the circle and in the flat piece of the
brane, (iii) scalar polarization which are not traceless and related
to radion field. We focus only on the transverse, traceless (TT) graviton 
which is represented by the metric perturbation around the classical 
solution of the 4D metric on the brane:  

\begin{equation}
ds^2 = e^{2f(r)} (\eta_{\mu\nu} + H_{\mu\nu}) dx^{\mu}dx^{\nu} + d r^2 +
e^{2 g(r)} L^2 d \theta^2  \label{eq:metricfluc}
\end{equation}

 The linearized gravitational fluctuation equation for the TT modes
can be written as  

\begin{equation}
\frac{1}{\sqrt{-g}} \partial_{M}(\sqrt{-g} g^{MN} \partial_{N}
H_{\mu\nu}) = 0 \label{eq:grav2}
\end{equation}

where, M, N denote the bulk spacetime indices.
For the above equation we can obtain solutions in the following form
$H_{\mu\nu} (x^{\mu},r,\theta) = h_{\mu\nu}(x^{\mu}) 
\sum_{lm} \varphi_{m}(r) e^{il\theta}$ 
where $h_{\mu\nu}$ satisfy the four dimensional field equation 
$\Box h_{\mu\nu} = m_0^2 h_{\mu\nu}$.  $m_{0}$ represents the mass 
of the corresponding modes. With these ansatze and in the background 
of the metric (\ref{eq:warp2}) the equation (\ref{eq:grav2}) reduces 
to the following form : 

\begin{equation}
\varphi_{m}'' + \left(\frac{m_0^2}{e^{\frac{k}{2} r} }  -
\frac{l^2}{L^2 e^{-2kr}} \right) \varphi_{m} = 0 
\end{equation}

where the modes $\varphi_{m}$ satisfy the orthonormality condition
 $2\pi \int_{0}^{\infty} dr L e^{2f + g}  \varphi_{m} \varphi_{n} =
\delta_{mn}$.  There exists a zero mass ($m_0 = 0$) and s - wave ($l = 0$) 
solution of the above equation given by $\varphi_{0} = Constant$. So
the normalized zero mode wave function can be written as 

\begin{equation}
\psi_{0} = \sqrt{\frac {k}{2 L}} e^{-\frac{k}{2} r}
\end{equation} 

which shows that the zero mode is localized near 
the origin $r = 0$. The modes for $m\neq 0$ and $l\neq 0$ may be obtained 
by solving the equation for $\varphi_m$ mentioned above. 
In (4 +$ p$) dimensions the modes will be localized 
for $p < 3$. 

In the case of the background metric with 
a bulk Brans-Dicke scalar, note that in the
Brans--Dicke gauge, the equations for the gravitational and Brans--Dicke 
scalar field fluctuations are $\Box H_{\mu\nu}=0$ and $\Box \xi =0$ ($\xi$
being the BD scalar fluctuation) \cite{linearbd}.  
Thus the BD scalar zero modes and the
graviton zero modes will be localized if the condition 
$(2k_{1} + k_{2}) < 0$ is obeyed. The fluctuation equations and hence the
condition for localization, will however change
provided we have other {\em matter} fields in the bulk.

\subsubsection{Scalar field (spin 0) :}

We now turn towards discussing the localization of 
a spin zero, scalar field in either of the fixed bulk background
line elements discussed in the previous subsection. 
The action for a massless scalar field coupled to gravity in D
dimensions is given by:

\begin{equation}
S_{0} = - \frac{1}{2} \int{d^{D} x \sqrt{-g} g^{AB} \partial_{A} \Phi
  \partial_{B} \Phi} \label{eq:scalaraction}
\end{equation}

where  A, B denote bulk spacetime indices. The localization condition
is equivalent to the normalizability of the ground state wave
function around the brane \cite{loc5}. The criterion will not be
different from the case of the graviton discussed in the previous section.
In a nutshell, one needs the 
integrals over the extra coordinates in the action to be finite so that the 
four dimensional part reduces to the usual 4D Klein-Gordon equation. 
In the background geometry given by the metric in (\ref{eq:metric}) 
the above action can be recast in the following form 

\begin{equation}
S_{0} = - \pi L \int_{0}^{\infty} {e^{2f(r) + g(r)} dr }
\int {\eta_{\mu\nu} \partial_{\mu} \phi \partial_{\nu} \phi d^{4} x}
\end{equation}
   

For the line element with the bulk phantom field, as given in equation 
(\ref{eq:warp2}) we see that the integral over $r$ is finite. So the 
scalar field zero modes will be localized on the brane.
For $ p$ extra dimensions one is restricted to $p < 3$ for the localization
of the zero mode.
In the case of bulk Brans-Dicke scalar the normalization
condition will be satisfied only if $(2 k_{1} + k_{2} < 0)$. The same
condition holds for six (4+p) dimensions as well.

\subsubsection{Vector field (spin 1) :}

It is known that in five dimensions the spin one vector (Maxwell) fields are 
not localized on the brane with increasing/decreasing  warp factors 
This is an inherent problem with the five dimensional models.
As we show below, in six
dimensions, the spin one fields can be localized on the brane. The action 
for a U(1) vector field reduces to the following form (with the choice of
a vector potential with no functional dependence on the extra
dimensional coordinates):  
\begin{eqnarray}
S_{1} &=& -\frac{1}{4} \int { d^{D} x \sqrt{-g} g^{AB} g^{MN} F_{AM}
  F_{BN}}  \\ \nonumber 
{} &=& - \frac{\pi}{2} L \int d r e^{g(r)} \int {d^{4} x \sqrt{-
  \eta} \eta^{\mu\nu} \eta^{\alpha\beta} F_{\alpha \mu} F_{\beta \nu}} 
\end{eqnarray}

If the integration over the extra coordinates is finite then
the above action will reduce to the standard Maxwell action in 
four dimensions.  
For the bulk phantom model g($r$) is a decaying function in six dimensions as
well as in (4+ $p$) - as a result the
integration over $r$ is finite and we can achieve localized U(1) gauge
field around the brane. 
In case of a bulk Brans-Dicke scalar the condition for the confinement 
of vector field on the brane is $k_{2} < 0$.


\subsubsection{Spinor field (spin $\frac{1}{2}$) :}

For spinor fields we need to look at the Dirac action in a 
D-dimensional warped spacetime. The Dirac equation is given by 

\begin{equation}
\Gamma^{A} V^{M}_{A} (\partial_{M} - \Omega_{M})
\Psi(x^{A}) = 0
\end{equation}
   
where, $V^{M}_{A}$ is the extension of the usual vierbein (tetrad) to 
six dimensions, $\Omega_{M} = \frac{1}{2}
\Omega_{M[AB]} \Sigma^{AB}$ the spin connection and $\Sigma^{AB} =
\frac{1}{4} \left[ \Gamma^{A},\Gamma^{B}\right]$, $\Gamma^{A}$ are 
the curved space gamma matrices. We look for
the solutions of the form $\Psi(x^{A}) = \psi(x^{\mu}) U(r)$ 
where $\Gamma^{\mu} \mathcal{D}_{\mu} = 0$ and $\mu$ stands for
brane coordinate index.
The assumption that the wave function does not
depend on $\theta$ leads to the solution $U(r) = u_{0} e^{-(2f + g/2)}$. 
Substituting this result into the action of the spinor fields
in curved spacetime we obtain

\begin{equation}
S_{Dirac} = u_{0}^2 \int e^{-f(r)}
dr \int {i \sqrt{-\eta} \bar{\psi} \gamma^{\mu} \partial_{\mu}
  \psi d^4 x} \label{eq:spinor1}
\end{equation}

The condition of trapping of spin $\frac{1}{2}$ fields on the brane
now becomes equivalent to having the integral over the extra coordinates 
as finite.
In Model-I  f($r$) is a growing function, so the integral is
finite and non-vanishing, which, in turn, guarantees the localization of
spin half fermions on the brane. In codimension p branes f($r$)
will remain an increasing function of $r$ only for $p > 1$.
To have localized fermions in Model-II one needs the condition $k_{1}
> 0$ to be satisfied for both the codimension two and codimension $p$ branes. 
This is easily satisfied by our models in both the cases. 

In summary, assuming that the standard model fields (eg. scalar, vector or
fermion) are independent of the extra dimensional coordinates, we find that
the criteria for existence of localized zero modes are related to the
finiteness of the following integrals:

\begin{equation}
\int_0^{\infty} e^{2f+g} dr ,\hspace{.1in} (scalar); \hspace{.1in}
\int_0^{\infty} e^{g} dr ,\hspace{.1in} (vector); \hspace{.1in}
\int_0^{\infty} e^{-f} dr ,\hspace{.1in} (fermion) \hspace{.1in}
\end{equation}

In addition, the localization criterion for the zero mode graviton 
is identical to that for the massless scalar.

\section{Brane models with conical singularity}

Till now, the models under consideration have been essentially
4--branes with a compact on--brane extra dimension. These models, therefore
represent a hybrid between the usual Kaluza--Klein idea
and the braneworld perspective where compactication is replaced by the
notion of localization. 
In this section, we consider the possibility of having 3-branes
as conical defects in the six dimensional spacetime.  
The metric for a general six
dimensional spacetime containing a warped
codimension two brane and obeying the four dimensional Poincare invariance
is given by 

\begin{equation}
ds^2 = g_{MN} dx^{M} dx^{N} = e^{2f(r)} \eta_{\mu\nu}dx^{\mu}dx^{\nu} + d r^2 +
e^{2 g(r)} d\theta^2  \label{eq:metric2}
\end{equation} 

The radial dimension is semi-infinite {\em{i.e.}} $0 \le r \le \infty$
and the coordinate $\theta$ is cyclic [$\theta : 0 \rightarrow 2
\pi$]. For the conical singularity at the location of the 
we impose the boundary
condition 

\begin{equation}
{e^{2 g(r)}}|_{r = 0} = 0 
\end{equation}

Note that if $e^{2g(r)}$ has multiple zeros (at different values of $r$),
it will be possible to place branes at those points in the extra dimensional
space.


The Einstein-scalar equations for a bulk phantom field with potential
V($\phi$) and a nonzero bulk cosmological constant $\Lambda$ lead to the
following solutions for the warp factors and the bulk scalar field 

\begin{eqnarray}
e^{2 f(r)} & = & \mbox{sech}^{2 \over 5} (kr) ~~~~~~~~~~~~~~~~~~~~
e^{2 g(r)}  =  \frac{\sinh^{2} (kr)}{\cosh^{2 \over 5} (kr)} \\
\phi(r) & = & \sqrt {\frac{4}{5 \alpha}} \ln \cosh(kr) \hspace{1.5 cm}
V(\phi) = - \frac{\Lambda}{\alpha} \tanh^{2} (kr)= \frac{\Lambda}{\alpha}
\left (e^{-\sqrt{\frac{5\alpha}{2}}
\phi}-1 \right )
\end{eqnarray}

where, $\alpha = {1 \over M^4}$ and k is an arbitrary constant. The bulk
cosmological constant is related to $k$ as $\Lambda = {{4 k^2} \over 5}$. 
From the above relations, it is clear that the brane warp factor
$e^{2f}$ decays as we go away from the brane location (at $r=0$).
The extra dimensional part of the line element has the factor
$e^{2g}$ which, as required, has a zero at $r=0$ and therefore
results in a conical deficit. The potential for the bulk scalar field
is negative definite for a positive bulk cosmological
constant. Another solution (with a growing warp factor) in vacuum has been 
obtained in {\cite{npb}}.

A natural question, following our previous discussion of 4--branes, is
--what happens if we have a Brans--Dicke scalar in the bulk?
To this end, we assume the following forms for $f(r)$, $g(r)$ and
$\ln\phi$:
\begin{eqnarray}
f(r) = \alpha \ln \cosh kr + \beta \ln \sinh kr \\
g(r) = \gamma \ln \cosh kr +\eta \ln \sinh kr \\
\ln \phi = \mu \ln \cosh kr + \nu \ln \sinh kr
\end{eqnarray}

Substituting the above ansatz in the Brans--Dicke equations mentioned
earlier and some manipulations, we end up with the following constraints :

\begin{equation}
\alpha +\beta = \gamma +\eta =0 \hspace{.1in}; \hspace{.1in}
\mu =-\left ( 4\alpha+\gamma -1 \right ) \hspace{.1in};
\hspace{.1in} \nu = -\left ( 4\beta + \eta -1 \right )
\end{equation}

An additional equation also exist which relates $\alpha$, $\gamma$ and
$\omega$.
These constraints show that the possible solutions for $f(r)$ and $g(r)$ are
necessarily singular (by virtue of the fact that the Ricci scalar will have a
divergence through the term proportional to $\coth^2 kr$).

Following a similar argument as in the case of 4--branes discussed earlier, 
and using the result
$\delta(\vec r) = \frac{1}{2\pi e^{g}} \delta(r)$ we find that
for the model with only a cosmological constant we get the brane tension
as $\lambda=-5\pi k$ whereas for the model we have derived with a
phantom bulk in a potential $\lambda=-2\pi k$.

Let us now provide a somewhat general method of constructing the
the warp factors $f$ and the extra dimensional factor $g(r)$
through the specification of a single function : the determinant of the
metric.
We first note that, for a minimally coupled bulk scalar or a bulk
phantom field dependent only on the transverse radial coordinate $r$, 
the Einstein tensor components $G_{00}$ and $G_{66}$ are related
by the equation, $G_{00} = - G_{66}$. This, in turn, leads to the 
opportunity to explore the spectrum of possibilities of having various 
solutions of the warp factors for different choices of the scalar potential.
The general solutions for $f(r)$ and $g(r)$ can be shown to 
functionally depend on the determinant ($\bar g$) of the bulk metric in the
following way:

\begin{eqnarray}
f(r) & = & \frac{1}{5} \ln \sqrt{- \bar{g}} - \frac{C}{5} \int \frac{d
  r'}{\sqrt{- \bar{g}}} + C_{1} \\
g(r) & = & \frac{1}{5} \ln \sqrt{- \bar{g}} + \frac{4 C}{5} \int
 \frac{d r'}{\sqrt{- \bar{g}}} + C_{1} 
\end{eqnarray}  
    
 where, C and $C_{1}$ are integration constants to be fixed by the
 other equations. In the following Table (I) we give examples of several
 toy models constructed by specifying a functional form for the
determinant. Note that to have a brane at a conical singularity
one must choose the determinant to have a zero at some specific point(s)
in the extra dimensional space. 

\begin{table}[htb]
\begin{center}
\begin{tabular}{|c|c|} \hline
 \multicolumn{2}{|c|} {Warp Factors} \\ \cline{1-2}
  $e^{2 f(r)}$& $e^{2 g(r)}$  \\ \hline
 $\mbox{sech}^{2 \over 5} (kr)$ & $\mbox{sinh}^{2}
  (kr){\mbox{cosh}^{-{2 \over 5}} (kr)}$  \\ \hline
 $ \mbox{e}^{- {2 \over 5} \mbox{sech}(kr)} R(r)
\mbox{cosh}^{4 \over 5} (kr / 2) $
& $\mbox{e}^{{8 \over 5} \mbox{sech}(kr)}  \frac{\mbox{cosh}^{4 \over 5}
(kr)}{ \mbox{cosh}^{6 \over 5} ({kr / 2}) } S(r)$ \\ \hline
 $\mbox{e}^{- {2 \over 5} \mbox{cosh}(kr)} \frac{\mbox{cosh}^{4 \over 5}
(kr/2)}{\mbox{cosh}^{4 \over 5}(kr)} $ &
$ \mbox{e}^{{8 \over 5} \mbox{cosh}(kr)}  \frac{\mbox{cosh}^{-{ 6 \over 5}}
(kr/2)}{ \mbox{cosh}^{4 \over 5}
(kr)} S(r) $  \\ \hline
 $ e^{-{2 \over 5} kr - {2 \over 5} \mbox{e}^{kr}}
\frac{\mbox{cosh}^{4 \over 5} (kr/2)}{\mbox{\mbox{cosh}}^{2 \over 5}(kr)} $ &
$\mbox{e}^{-{2 \over 5} kr+ {8 \over 5} \mbox{e}^{kr}}
\frac{\mbox{cosh}^{-{ 6 \over 5}} (kr/2)}{ \mbox{cosh}^{2 \over
5} (kr)} S(r)$   \\ \hline
\end{tabular}
\caption{$R(r) =
\mbox{cosh}^{4 \over 5} (kr)$, $S(r)$ =  $\mbox{sinh}^{2} (kr/2)$.
The solutions for the warp factors are given for four
different choices of the determinant of the bulk metric $\bar g =
\mbox{tanh}(kr),\mbox{sinh}(kr) \mbox{cosh}^2 (kr)$, $\mbox{sech}(kr)
\mbox{tanh}(kr)$ and $e^{- kr} \mbox{tanh}(kr)$ respectively.}
\end{center}
\end{table}

The full solution of the scalar field equation has been already 
discussed for the first example but it is not easy to 
obtain the exact solutions for the scalar field in the other three 
cases. It can be shown, however, from a graphical analysis of the variation 
of $\phi'^2$ w.r.t. radial coordinate r that the scalar field is real
everywhere. 
The potential can be represented as a function of r in all the three cases
though its representation as a function of $\phi$ depends on the
analytical solvability of the scalar field equation.
It must be admitted that
the above models are all pretty complicated and contrived in nature.
We do not intend to discuss these models any further here.

\subsection{Localization of gravity and matter fields on the brane
  with a conical singularity}

Let us now discuss the localization scenario for the 3-brane
models discussed above. The transverse traceless
modes of the linearized gravity fluctuation polarized along Lorentz invariant
hypersurface satisfies the equation given in (\ref{eq:grav2}). For an
infinitely extended transverse radial dimension we don't find any
normalizable zero mode for the models given in the Table (I)
because the normalization integral $\int_{0}^{\infty} e^{2f(r)
  + g(r)} dr$ is not finite in any of the cases. As a remedy,
normalized zero modes may be achieved by truncating the radial
direction at a certain value of r and placing a 4-brane (with a
compact on--brane extra dimension) at that
point \cite{kogan}. The picture then will be like the {\sc rs-i}
model and the gravity localization problem will be similar to the
corresponding Schr\"{o}dinger problem. Along with the
localized massless mode one will also find discrete Kaluza-Klein
modes. As we have discussed earlier, the same conditions are
also applicable for the confinement of a scalar field. In the background
geometry described by the warp factors in the second and fourth rows of Table (I) we
find localized spin 1/2 zero modes on the three-brane. The integration over
the transverse radial coordinate in equation (\ref{eq:spinor1}) is finite in
these two cases. It turns out that the spin-1 vector fields also do not have
any massless mode localized on the 3-brane.

One may ask, what could be the functional forms of $f(r)$ and $g(r)$ in order
to have localization of all the fields? Let us go back to the ansatz we had
made earlier:

\begin{eqnarray}
f(r) = \alpha \ln \cosh kr + \beta \ln \sinh kr \\
g(r) = \gamma \ln \cosh kr +\eta\ln \sinh kr
\end{eqnarray}

Obviously, we do not want the warp factor ($e^{2f}$) to go to either zero
or infinity at the location of the brane. This implies we must have
$\beta =0$. Also, to have a zero in $e^{2g(r)}$ at the location of the
brane, we need $\eta >0$. With these choices, we can now go back to the
requirement that the three integrals $\int e^{2f+g} dr$, $\int e^{-f} dr$
and $\int e^{g}dr$ should give finite answers. The divergence of these
integrals will stem from large values of r (i.e. $r\rightarrow \infty$).
Taking the asymptotic forms of $\cosh kr$ and $\sinh kr$ as $r\rightarrow
\infty$, which, essentially implies including only the exponentially
growing pieces in each of these, we find that the finiteness of the
integrals imply that the following holds:

\begin{equation}
2\alpha + \gamma + \eta < 0 \hspace{.1in};\hspace{.1in} \alpha >0 \hspace{.1in};
\hspace{.1in} \gamma +\eta <0
\end{equation}

Using $\alpha >0$, we find we must have $\gamma  <-2 \alpha-\eta$.
It can be shown easily that the two solutions we have quoted earlier
do not satisfy these constraints. If we can have a solution of the
six dimensional field equations with the coefficients $\alpha$, $\gamma$
and $\eta$ satisfying the above mentioned constraints we will have a
warped codimension two braneworld located at a conical singularity in the
6D bulk,  where all the above three fields will be  localized.

\subsection{\bf Localizing all fields: geometries and energy conditions}

From the above discussion, we can write down, in an ad-hoc manner,
warped geometries with a conical singularity for which the zero modes
of all fields will be localized. In this brief section, we provide
some examples and then, investigate, the nature of matter through
an analysis of some of the energy conditions \cite{visserbook}. We do not attempt here
to arrive at {\em solutions} to the Einstein equations in the presence
of specified matter sources (such as a scalar field etc.). Our aim is
to see whether we can have line elements which are non--singular and with
properties enabling it to represent a warped braneworld in six dimensions
where the brane is located at a conical singularity in the bulk.

Recall the forms of $f$, $g$ mentioned in the previous section. We
assume $\eta=1$ -- this avoids the presence of a singularity at $r=0$.
We also keep in mind the various constraints on $\alpha$ and $\gamma$.

The various components of the energy momentum tensor turn out to be :
\begin{eqnarray}
\rho & = & \kappa \left [ (-6\alpha^2+3\alpha-3\alpha\gamma +\gamma -\gamma^2)\tanh^2 kr
+(-6\alpha - 3\gamma -1 ) \right ]\\
p_{x,y,z} & =& \kappa \left [ (6\alpha^2-3\alpha+3\alpha\gamma -\gamma +\gamma^2)\tanh^2 kr
+(6\alpha + 3\gamma +1 ) \right ]\\
p_r &=& \kappa \left [ (6\alpha^2 + 4\alpha\gamma ) \tanh^2 kr + 4\alpha \right ]
\\
p_\theta &=& \kappa \left [ (10\alpha^2-4\alpha ) \tanh^2 kr + 4\alpha \right ]
\end{eqnarray}

One can consider looking at the inequalities that need to hold if the
Null Energy Condition ( $\rho+p_i \ge 0$) or the Weak Energy Condition
($\rho \ge 0$, $\rho+p_i \ge 0$ ) have to be satisfied.
It is easy to note that $\rho+p_{x,y,z}=0$ but the other inequalities
need to be checked. It turns out that the requirement $2\alpha+\gamma+1\le 0$
is not compatible with what one needs in order to satisfy the energy conditions.
In other words, violation seems to be a necessity in order to achieve localization.

As an example, we choose $\alpha =\frac{1}{2}$, $\gamma = - 3$ and write down
the various components for this choice.
\begin{eqnarray}
\rho& =& - \kappa \left ( \frac{15}{2} \tanh^2 kr - 5 \right ) = -p_{x,y,z} \\
p_r &= & \kappa \left (-\frac{9}{2} \tanh^2 kr + 2 \right );~~ p_\theta =
\kappa \left ( \frac{1}{2} \tanh^2 kr +2 \right ) 
\end{eqnarray}

where a $k^2$ has been absorbed in the definition of $\kappa$.

Notice, even though it is negative the energy density is bounded. So, are the
pressures. The line element is :
\begin{equation}
ds^2 = \cosh kr \left ( \eta_{\mu\nu}dx^\mu dx^{\nu} \right ) + dr^2 +
\frac{\sinh^2 kr}{\cosh^6 kr} d\theta^2
\end{equation}

The above is an example of a warped six dimensional line element for which all
fields will be localized on the brane and the brane will be located at
a conical singularity in the bulk line element.
This, of course is not the only possibility, there are infinitely many
such line elements. It will be nice to find bulk matter sources which can
generate them.

\section{SUMMARY AND CONCLUSIONS}

Let us now summarize the results obtained in this paper and discuss 
the open issues.

Exact bulk solutions for some new, six dimensional braneworld  models are 
obtained for the bulk phantom scalar field and the Brans-Dicke scalar field respectively. 
We have shown that the phantom field and the BD scalar 
can provide the adequate source terms (in the Einstein equations) which 
enable the existence of various types of solutions. In particular, for the
BD scalar we find that there are solutions which can have a decaying warp
factor and a growing extra dimensional factor and vice versa.  
The above mentioned class of models are essentially 4--brane models
with an on--brane, compact (angular) extra dimension. Further, in these models,
we address the issue of localization of gravity as well
as other fields in this context.
A unique feature of the models is the
localization of massless spin fields ranging from 0 to 2 on a 
single brane by means of gravity only. 
In particular, the sixth dimension seems to facilitate the localization of
vector fields, a result which does not exist in five dimensions.
We have also generalized the results 
obtained for codimension two branes to
that for codimension $p$ branes with ($p$ - 1) compact ($S^1$) and one
non -- compact extra dimension.
Subsequently, we have studied the genuine 3--brane models where the 
brane is located at a conical singularity in the bulk. In the presence of a 
bulk phantom scalar  we have constructed a viable model with a 3--brane.
We then discuss a general method of constructing the warp factor and the
extra dimensional factor with the determinant of the bulk metric as the only 
input. Finally, we address the localization scenario using the generic criteria
obtained in the previous section. Apart from localized fermions, none of the
other fields can be localized within this class of models with the brane as 
a conical defect.  
We point out (with examples) the conditions under which, for a sufficiently
broad class of warp and extra dimensional factors, we can construct
a model with a conical singularity where all the massless modes of the
standard model fields can be localized. 

The energy-momentum tensor corresponding to the bulk phantom scalar
violates all the energy conditions and the bulk spacetime obtained 
in this setup is probably not dynamically stable. However, in the bulk, 
the brane is geometrically stable against small normal deformations (analysed
via the Jacobi equations 
\cite{stab}) in a {\em fixed} bulk. 
In this regard, we mention that the bulk in the RS model \cite{rs2} also
violates the energy conditions but the brane is not geometrically  
stable under small normal deformations in a fixed bulk \cite{stab}. 
In order to investigate the full dynamical 
stability of the bulk, we need to look at the full gravitational as well as 
scalar field fluctuations (perturbing both sides of the Einstein equations)
and obtain the resulting criteria. Note that there
are possible subtleties which may arise because of the presence of the
two functions $f$ and $g$ in the bulk metric, which can have growing/decaying
characteristics. Merely discarding these models as unstable because of the 
phantom's presence is possibly not the right thing to do {\cite{peloso}}.
We intend to give this issue a more careful look later.       

A further point concerns localization issues for the bulk solution
with a Brans--Dicke field. We have stated that the criteria for
gravity localization is the same as for the phantom (modulo the
nature and coefficients in the warp factor). This happens because
the equations for the perturbations of the bulk metric and the Brans--Dicke
scalar in the so--called Brans--Dicke gauge remain the same as that for the
bulk phantom. If one
places extra matter sources on the brane or in the bulk, things will
surely change (because the perturbation equations change). 
It will be interesting to probe, what kind of gravity
a Brans--Dicke bulk can induce on the brane. Recall that in the work of
Garriga--Tanaka \cite{gt}, the gravity induced on the positive and negative
tension branes were of Brans--Dicke type (with different Brans--Dicke
parameters).  

Additionally, we need to reconsider issues like solution of
hierarchy problem, construction of Friedman--Robertson--Walker (FRW) 
branes in a six--
dimensional bulk, the stability of such FRW  branes under fluctuations 
and graviton massive mode localization in the background geometries 
describing 3 and 4-branes. It also remains an open issue to find 
a {\em good} model with a conical singularity for a given matter source (eg.
scalar fields with some potential) and with 
all the known matter fields localized. 

We hope to make progress with the above aspects in our forthcoming articles.


\section*{Acknowledgments}

RK thanks CSIR, India for financial support through a Senior Research
Fellowship.


\end{document}